\documentclass[%
 reprint,
 amsmath,amssymb,
 aps,
]{revtex4-2}

\usepackage{graphicx}
\usepackage{dcolumn}
\usepackage{bm}
\usepackage{physics}
\usepackage{mathtools}
\usepackage{float}
\usepackage[usenames,dvipsnames]{color}
\usepackage[colorlinks=false, pdfborder={0 0 1}]{hyperref}

\begin{document}

\preprint{APS/123-QED}

\title{Tailoring Spatial Modes Produced \\by Stimulated Parametric Downconversion}

\author{A.A. Aguilar-Cardoso}
\email{aagui026@uottawa.ca}
\author{C. Li}%
\author{T.J.B. Luck}
\author{M.F. Ferrer-Garcia}
\author{J. Upham}

\affiliation{
Department of Physics, University of Ottawa, 25 Templeton, Ottawa,
 K1N 6N5, ON, Canada.
}%

\author{J.S. Lundeen}%
\affiliation{
Department of Physics, University of Ottawa, 25 Templeton, Ottawa,
 K1N 6N5, ON, Canada.
}%
\affiliation{
 National Research Council of Canada, 100 Sussex Drive, Ottawa, Ontario K1A 0R6, Canada.
}%
\author{R.W. Boyd}
\affiliation{
Department of Physics, University of Ottawa, 25 Templeton, Ottawa,
 K1N 6N5, ON, Canada.
}%
\affiliation{
Institute of Optics, University of Rochester, Rochester, 14627, NY, USA.
}%

\date{\today}

\begin{abstract}
We theoretically study and experimentally demonstrate controlled generation of spatial modes of light via stimulated parametric downconversion (StimPDC) by transferring the spatial structure of a pump beam to the stimulated idler beam. We show how the beam characteristics of the stimulated beam depends on both the pump and seed beam's characteristics, enabling experimental control over size and propagation behavior. We also show how to control and improve the fidelity of different spatial modes, and demonstrate that the angular basis ensures uniform fidelity across modes generated with StimPDC.
\end{abstract}

\maketitle


\section{Introduction}

Spatial modes of light, optical fields characterized by a transverse spatial distribution that is maintained during free-space propagation, has proven useful in a wide range of fields such as telecommunications~\cite{BOUCHARD-2018,klug-2023}, microscopy~\cite{boyd:20}, and machine learning \cite{Fang:24}. Typically, these structured optical fields are generated by linear optical methods including digital holography using  spatial light modulators (SLMs)~\cite{rosales2017shape}, liquid-crystal-based q-plates ~\cite{rubano2019q}, and engineered metasurfaces \cite{dorrah2022tunable}. Recently, there has been a growing interest in harnessing nonlinear optical (NLO) processes to generate and manipulate structured light. Unlike linear methods, NLO interactions enable in-situ tailoring of spatial and spectral properties during propagation or frequency conversion, offering dynamic control and access to novel regimes of light–matter interaction~\cite{buono-2022,jones2024ultrafast}. 

\begin{figure}[ht]
    \centering
    \includegraphics[width=0.8\linewidth]{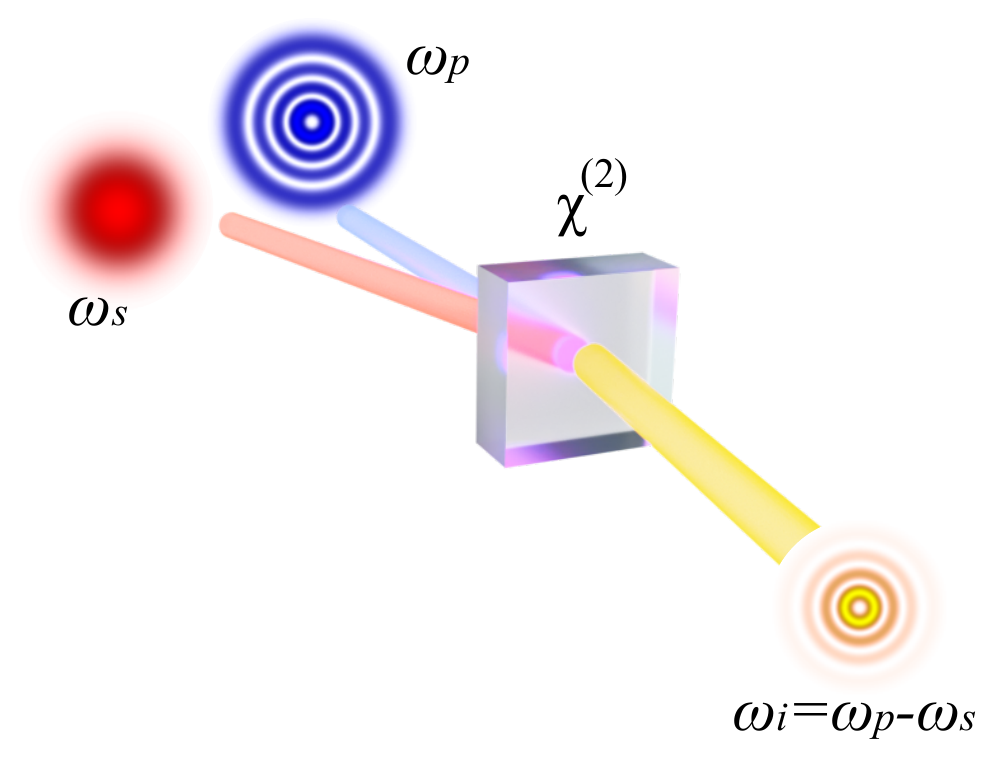}
    \caption{Illustration of structured light generation using StimPDC. A structured pump beam at frequency $\omega_p$ interacts with a seed beam at $\omega_s$, stimulating the emission of an idler beam at $\omega_i = \omega_p - \omega_s$ that inherits the spatial structure of the pump.}
    \label{fig:intro}
\end{figure}

A particularly promising NLO process for structured light generation is stimulated parametric downconversion (StimPDC). In StimPDC, a nonlinear crystal is injected with a pump beam at frequency $\omega_p$ and a seed beam with frequency $\omega_s$ to produce a stimulated beam at frequency $\omega_i=\omega_p-\omega_s$. The spatial counterpart of this frequency relation, which is governed by the nonlinear phase-matching parameters, allows conditioning of the spatial structures of the stimulated beam on those of the pump and seed beams, as illustrated in Fig. \ref{fig:intro}. For instance, StimPDC can be tailored to produce an idler beam with the phase conjugation of the seed beam ~\cite{de-oliveira-2019,dos2022phase}, enabling a variety of applications such as the efficient generation and detection of quantum states~\cite{Liscidini:13,fadrny-2024,xu-2024} and information encoding~\cite{Xu:23}. Although not the focus of the current work, the polarization state of this simulated beam can be engineered through the optical properties of the nonlinear medium, allowing the generation of beams with spatially varying polarization \cite{de-oliveira-2020}.

While recent work has shown that the fidelity of structured light generated in the stimulated field is governed by the effective transverse overlap between the pump and seed beams~\cite{singh-2025}, a comprehensive study that quantifies the fidelity of spatial mode generation via StimPDC and their longitudinal propagation is essential to assess its viability for the aforementioned advanced photonic applications. In this work, we analyze the joint effect of the seed and pump beams' transverse structures and propagation properties on those of the generated stimulated beam. This paper is organized as follows. Section \ref{Tailor} presents a theoretical framework on how the seed and pump beams' spatial dimensions affect the stimulated beam's Rayleigh range, i.e. the field's propagation divergence. Section \ref{spatial_modes} evaluates fidelity between the target spatial mode and the output field generated through the nonlinear interaction. In Section \ref{experiments}, we experimentally demonstrate the transfer of spatial structures in StimPDC and show that our theoretical predictions apply for a variety of spatial mode bases, such as Laguerre-Gauss (LG), Hermite-Gauss (HG), and angular (ANG) modes. We believe that these findings contribute to understanding the nonlinear optical methods for generating structured light and highlight the potential of StimPDC as a platform for high-fidelity mode transfer in optical communication systems. 

\section{Tailoring StimPDC}
\label{Tailor}

We assume that the seed beam is sufficiently strong such that the contribution from spontaneous parametric downconversion (SPDC) is negligible compared to that of the StimPDC. This allows modeling of the interaction within the classical regime of nonlinear optics~\cite{arruda-2018}. In addition, we consider that the nonlinear interaction between the complex field of the pump beam $U_p$ and the seed beam $U_s$ occurs within a nonlinear crystal whose thickness ($L$) is much smaller than the Rayleigh range of the incident beams ($z_{Rp}$,$z_{Rs}$). This thin-crystal approximation allows us to disregard the propagation effects induced within the nonlinear medium, reducing the interaction to a phase-matching condition applied at a single transverse plane. Under this framework, it is possible to approximately describe the spatial profile of the generated stimulated beam $U_i(\mathbf{r})$ as \cite{Boyd:2020}
\begin{equation}
\label{nolinear}
    U_{i}(r)\propto U_p(r,0)U_s^*(r,0),
\end{equation}
where the subscripts $p$, $s $ and $i$ identify the pump, seed and stimulated beams, respectively, and $r$ is the transverse spatial coordinates of the field.

The simulated beam generated from nonlinear interactions have free-space propagation properties that differ significantly from both the pump and seed beams. To describe this effect, let us consider that both the pump and the seed beams have a Gaussian profile with waist parameters $w_p\neq w_s$ at the center of the nonlinear crystal. The normalized complex amplitude of a Gaussian beam in cylindrical coordinates $(r,\, \phi, \, z)$ is given by

\begin{equation}
\label{gauss}
    U(r,z)=\frac{U_0}{w(z)}e^{-\frac{r^2}{w^2(z)}}e^{i\left(\frac{k r^2}{2R(z)}-\Phi(z)\right)},
\end{equation}
where $z$ is the central propagation direction of the beam, $\Phi(z)$ and $R(z)$ are the Gouy phase and radius of curvature of the wavefront at $z$, $k=2 \pi/ \lambda$ is the wavenumber, and $\lambda$ is the wavelength of interest. The beam waist evolves along propagation according to $w(z)=w_0 \sqrt{1+\left(\frac{z}{z_\textit{R}}\right)^2}$, where $w_0$ is the beam waist at the plane $z=0$ and $z_\textit{R}=\frac{\pi w_0^2}{\lambda}$ is the Rayleigh range. Eq. \ref{gauss} is solution to the paraxial wave equation and it represents the fundamental spatial mode \cite{saleh-1991}. 

\begin{figure}[ht]
    \centering
    \includegraphics[width=0.8\linewidth]{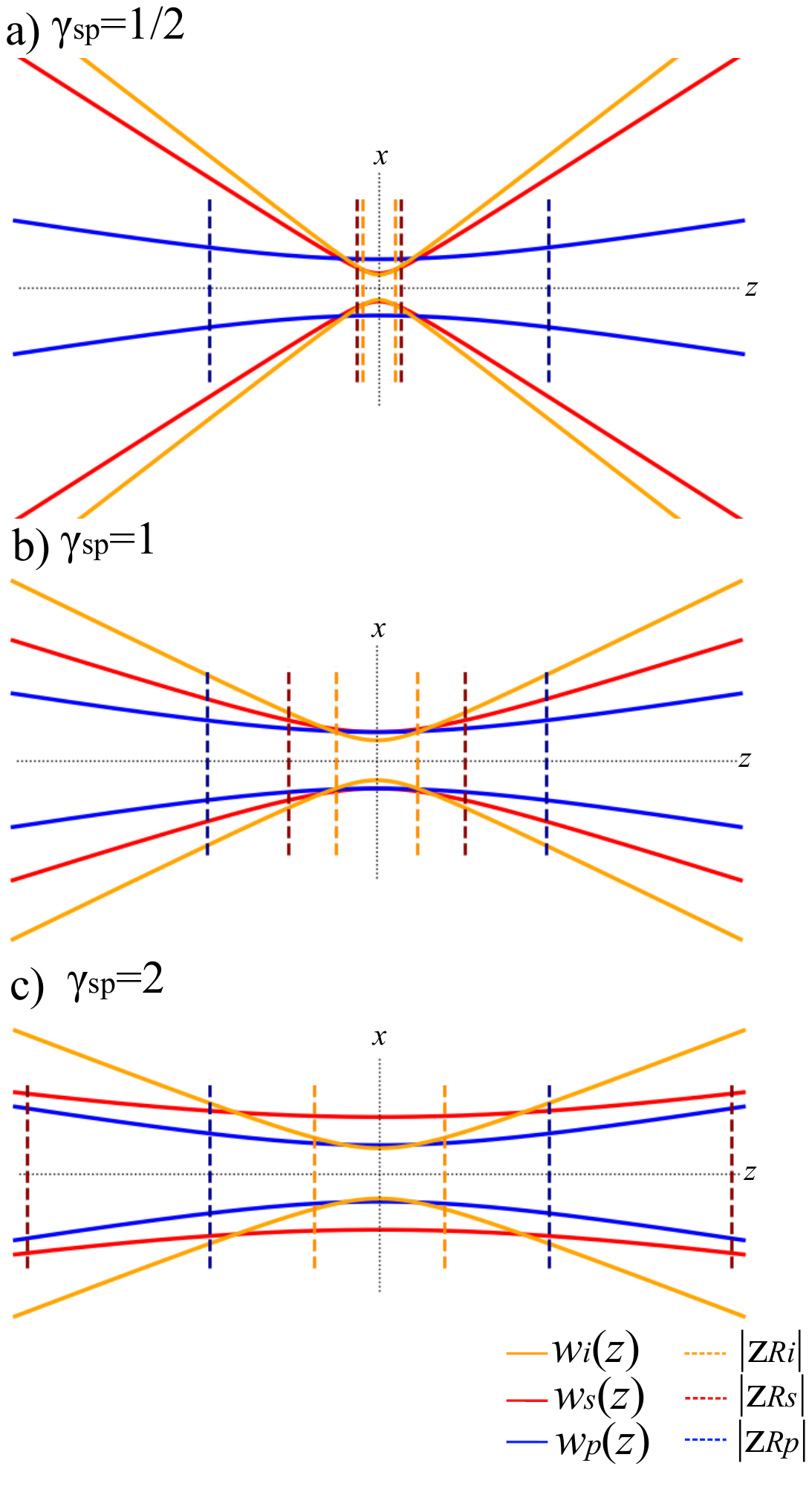}
    \caption{Beam waist propagation of pump beam at $\lambda_p=$ 405 nm (blue), seed beam at $\lambda_s=$ 780 nm (red) and stimulated beam at $\lambda_i=$ 842 nm (orange) for \textbf{a)} $\gamma_\textit{sp}=1/2$, \textbf{b)} $\gamma_\textit{sp}=1$ and \textbf{c)} $\gamma_\textit{sp}=2$. As $\gamma_\textit{sp}$ increases, the stimulated beam waist $w_s$ approaches that of the pump beam $w_p$, while its Rayleigh range converges to approximately half of the pump beam's Rayleigh range.
}
    \label{fig:plots}
\end{figure}

By substituting Eq. \ref{gauss} in Eq. \ref{nolinear}, it is easy to show that the beam waist parameters of the pump $w_p$, seed $w_s$, and stimulated beam $w_i$  are related by

\begin{subequations}
\label{waist}
\begin{align}
\label{waists}
\frac{w_i}{w_s}=\frac{1}{\sqrt{1+\gamma_\textit{sp}^2}},\\
\label{waistp}
\frac{w_i}{w_p}=\frac{1}{\sqrt{1+\frac{1}{\gamma_\textit{sp}^2}}},
\end{align}
\end{subequations}
where we have defined $\gamma_\textit{sp}={w_s}/{w_p}$. As shown in Eqs. \ref{waist}, the beam waist of the output beam $w_i$ solely depends on the ratio of the waists of the pump and seed beams. These two ratios serve as key factors for controlling the structure of the generated field. Fig. \ref{fig:plots} depicts the effect of varying $\gamma_\textit{sp}$ when fixing the ratio between wavelengths of the pump and seed beams. For example, when $\gamma_\textit{sp} \ll 1$, the pump field appears relatively constant over the transverse profile of the seed beam, which results in $w_i \approx w_s$. On the other hand, as the value of $\gamma_\textit{sp}$ increases, $w_i$ asymptotically approaches $w_p$. Finally, for $\gamma_\textit{sp} = 1$, where both pump and seed beams have equal waists, the waist of the stimulated beam yields $w_i =  w_{p,s}/\sqrt{2} $. Intuitively, one can understand the above behavior by treating the waist ratio $\gamma_\textit{sp}$ as a metric for describing the effective nonlinear interaction region.

Using these results for the waist of the output beam, we proceed to analyze the propagation of the output beam. It is possible to write the Rayleigh range of the stimulated beam in terms of the ratios,
\begin{subequations}
\label{range}
\begin{align}
\label{ranges}
\frac{z_\textit{Ri}}{z_\textit{Rs}}&=\left(\frac{\lambda_s}{\lambda_p}-1 \right) \left(\frac{1}{1+\gamma_\textit{sp}^2} \right),\\
\label{rangep}
\frac{z_\textit{Ri}}{z_\textit{Rp}}&=\left(1 - \frac{\lambda_p}{\lambda_s} \right) \left(\frac{1}{1+\frac{1}{\gamma_\textit{sp}^2}}\right),
\end{align}
\end{subequations}
where $\lambda_p <\lambda_s$. Notice that the Rayleigh range of the stimulated beam depends exclusively on the waist ratio $\gamma_\textit{sp}$ and the wavelength ratio $\lambda_s/\lambda_p$ of the seed and pump beams. As depicted in Fig. ~\ref{fig:plots}a, for very small values of $\gamma_\textit{sp}$, $z_\textit{Ri}$ becomes comparable to that of the seed. Meawhile, as depicted in Fig. \ref{fig:plots}b and \ref{fig:plots}c, $z_\textit{Ri}$ becomes smaller than $z_\textit{Rp}$ for larger values of $\gamma_\textit{sp}$, where the maximum value of $z_\textit{Ri}=z_\textit{Rp}/2$ is achieved for the particular case of degenerate StimPDC ($\lambda_s=\lambda_i$). Similarly, although the wavelength ratio $\lambda_s/\lambda_p$ is constrained by both the phase-matching requirements of the nonlinear crystal and the wavelength availability of laser sources~\cite{Boyd:2020}, they offer an additional degree of control over the stimulated beam. Our results could serve as a general guidance for designing a StimPDC-based light structuring system.

Although our analysis has only considered seed and pump beams with Gaussian profiles, we note that all previously discussed results regarding beam waist and divergence remain valid for any solution of the paraxial wave equation. To analyze the generation of more complex spatial modes, it is necessary to define a measure that quantifies the spatial mode transfer from the pump beam to the stimulated idler beam.
\section{Transferring spatial modes with StimPDC}
\label{spatial_modes}

In order to define the quality of the spatial mode inherited by the stimulated beam, we calculate the fidelity between the encoded spatial mode on the pump beam $U_p(\mathbf{r})$ and the resulting spatial mode on the stimulated field $U_i(\mathbf{r})$ as 

\begin{equation}
\label{fide}
    F=\frac{\left|\int U_{p}^*(\mathbf{r}) U_{i}(\mathbf{r}) d\mathbf{r}\right|^2}{\int U_{p}^*(\mathbf{r}) U_{p}(\mathbf{r}) d\mathbf{r}\int U_{i}^*(\mathbf{r}) U_{i}(\mathbf{r}) d\mathbf{r}},
\end{equation}
where $F=1$ corresponds to the case when both the spatial profile of the pump and stimulated beams completely overlaps, meaning the pump's spatial mode was perfectly copied to the idler beam.

Since the spatial profile of the pump and seed beams are related to the stimulated beam by Eq. \ref{nolinear}, we analytically calculate the fidelity as a function of $\gamma_\textit{sp}$ for each spatial mode. To simplify the subsequent analysis, we set a non-trivial complex amplitude profile for the pump beam, while keeping a Gaussian profile for the seed beam. 

\begin{center}
\begin{figure}[ht]
\includegraphics[width=0.8\linewidth]{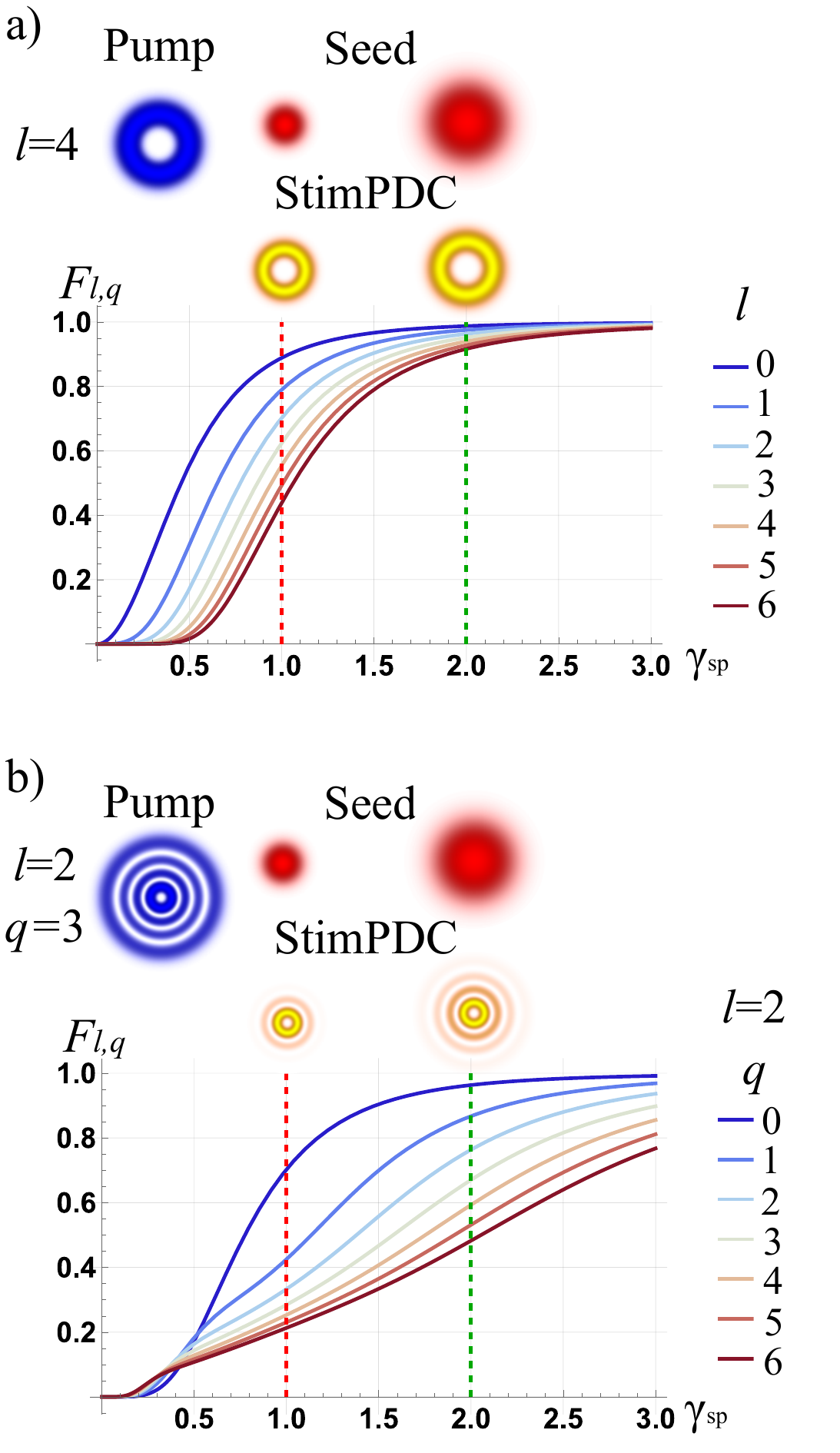}
\caption{\label{fig:OAM} Fidelity plots of stimulated beams calculated using Eq. \ref{fidLG} for different modal numbers as function of $\gamma_\textit{sp}$. (a) OAM beams ($q=0$) for $l\in[-6,6]$. (b) LG beams with $l=2$ and $q\in[0,6]$. The insets show examples of the transverse intensity profiles for both cases.}
\end{figure}
\end{center}

We start by considering the case that $U_p(\mathbf{r})$ is in an LG mode. The normalized complex amplitude of an LG beam in cylindrical coordinates $(r,\theta,z)$ is written as \cite{saleh-1991}, 

\begin{gather}
    \text{LG}_q^l(r,z)=\frac{\mathcal{N}}{w(z)}\left(\frac{\sqrt{2}r}{w(z)}\right)^{|\ell|}L_q^{|\ell|}\left(\frac{2r^2}{w^2(z)}\right)\nonumber\\  \times e^{-\frac{r^2}{w^2(z)}}e^{i\left(\ell\theta+\frac{k r^2}{2R(z)}-\Phi(z)\right)},
    \label{LG}
\end{gather}
where $L_{q}^{|\ell|}(\cdot)$ represents the generalized Laguerre polynomial, and the functions $R(z)$, $\Phi(z)$, and $w(z)$ retain the same definitions as in Eq. \ref{gauss}. The azimuthal index $\ell$, which is also referred to as topological charge, determines the number and direction of $2\pi$ phase cycles around the optical axis. The radial index $q$ indicates that the transverse intensity profile contains $q+1$ concentric rings. Substituting \ref{LG} into \ref{fide}, we can calculate the fidelity as
\begin{gather}
    F_{l,q}(\gamma_\textit{sp})=\nonumber\\\frac{\left(\int_0^\infty e^{- \frac{u}{2\gamma_\textit{sp}^2}} e^{-u } u^{|l|} [L_q^{|l|} (u)]^2 du\right)^2}{\int_0^\infty e^{-u } u^{|l|} [L_q^{|l|} (u)]^2 du \int_0^\infty e^{- \frac{u}{\gamma_\textit{sp}^2}} e^{-u } u^{|l|} [L_q^{|l|} (u)]^2 du},
    \label{fidLG}
\end{gather}

where $u=2r^2/w_p^2$, and the definition $\gamma_\textit{sp}={w_s}/{w_p}$ still holds for any spatial mode. Fig. \ref{fig:OAM} shows the solutions of the fidelity integral, given by equation \ref{fidLG}, for different modal numbers. Note that for a fixed value of the beam waist defined in Eq. \ref{LG}, the size of the transverse spatial profile increases with modal numbers. Consequently, the overlap between the LG structure and the seed beam becomes progressively smaller for higher-order beams. This results in a decrease in the fidelity of the stimulated beam as $l$ and $q$ increases. This occurs because at large $\gamma_\textit{sp}$, the effective interaction region is much smaller than the mode profile of the seed beam, such that the wavefront of the seed beam approximates that of a plane wave. As a result, the stimulated field completely inherits the structure of the pump field. This behavior is also evident in Eq. \ref{nolinear}, where, as the seed beam waist approaches infinity, the seed field's spatial structure becomes indiscernible in the nonlinear interaction region, resulting in $U_{i}(r)\propto U_p(r,0)$. 

Similarly, we consider the case of a pump beam described by an HG mode. The normalized amplitude of an HG beam is written as  \cite{saleh-1991},

\begin{gather}
    \text{HG}_{n_x,n_y}(r,z)=\frac{\mathcal{N}}{w(z)}H_{n_x}\left( 
 \frac{\sqrt{2}x}{w(z)}\right) H_{n_y}\left( 
 \frac{\sqrt{2}y}{w(z)}\right)\nonumber \\\times e^{-\frac{r^2}{w^2(z)}}e^{i\left(\frac{k r^2}{2R(z)}-\Phi(z)\right)},
 \label{HG}
\end{gather}
where $r^2=x^2+y^2$. $H_{m}(\cdot)$ is the generalized Hermite polynomials of index $m$, and the functions $R(z)$, $\Phi(z)$, and $w(z)$ retain the same definitions as in Eq. \ref{gauss}. Their transverse intensity consists of $(n_x+1)(n_y+1)$ bright lobes arranged in a rectangular grid, with nodal lines separating adjacent regions along the horizontal and vertical axes. Since the expression for an HG mode can be factorized into a product of its orthogonal transverse components, we can write the fidelity as
\begin{equation}
\label{fidHG}
    F_{n_x,n_y}(\gamma_\textit{sp})=F_{n_x}(\gamma_\textit{sp})F_{n_y}(\gamma_\textit{sp}),
\end{equation}
where $F_{n_{x(y)}}(\gamma_\textit{sp})$ denotes the fidelity calculated for the x(y)-components and is explicitly written as

\begin{gather}
    F_{n_j}(\gamma_\textit{sp})=\nonumber\\\frac{\left( \int_{-\infty}^\infty e^{- \frac{u^2}{2\gamma_\textit{sp}^2}} e^{-u^2 } [H_{n_j} (u)]^2 du \right)^2}{   \int_{-\infty}^\infty  e^{-u^2 } [H_{n_j} (u)]^2 du  \int_{-\infty}^\infty e^{- \frac{u^2}{\gamma_\textit{sp}^2}} e^{-u^2 } [H_{n_j} (u)]^2 du     }.
\end{gather}

\begin{figure}[ht]
    \centering
    \includegraphics[width=0.8\linewidth]{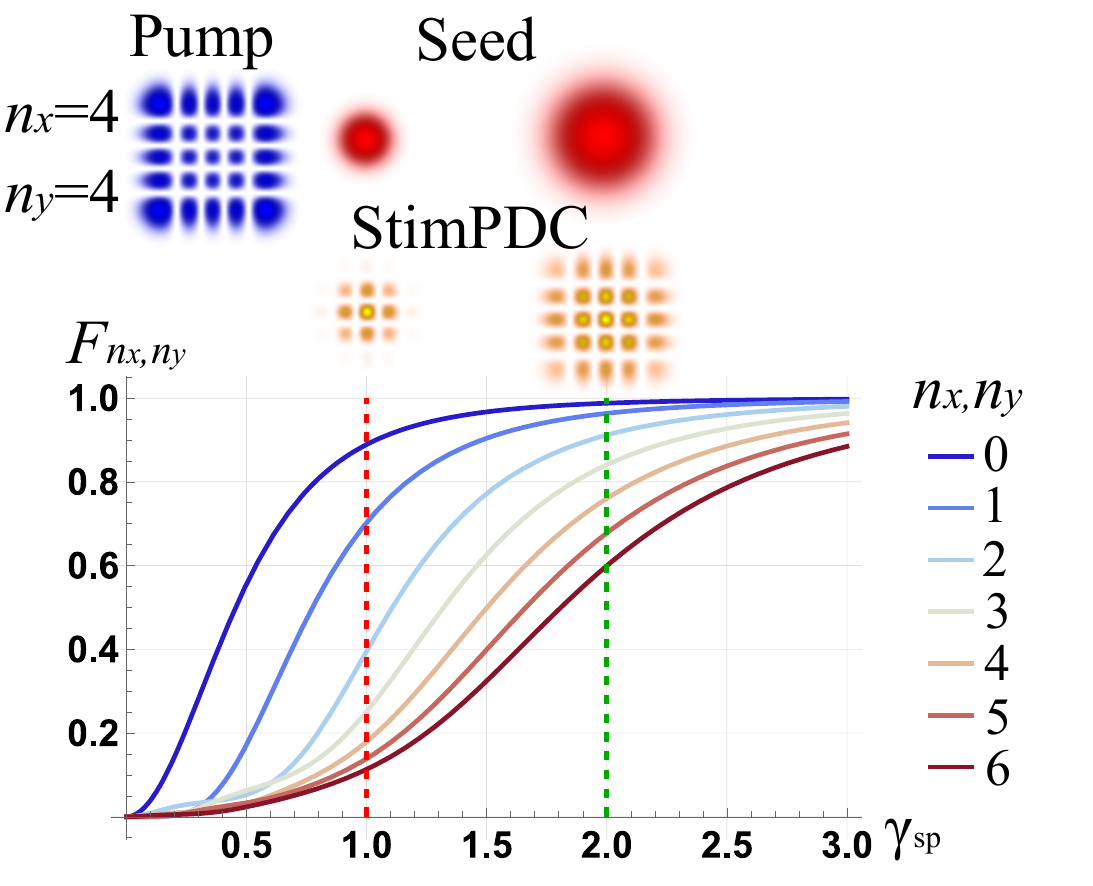}
    \caption{Fidelity plots of stimulated beams calculated using Eq. \ref{fidHG} for HG beams with $ n_x = n_y \in[0,6]$ as function of $\gamma_\textit{sp}$. The insets show examples of the transverse intensity profiles.}
    \label{fig:HG}
\end{figure}

Fig. \ref{fig:HG}b shows the fidelity values for HG modes of different modal numbers. We notice a similar behavior to the case of LG modes. Since the number of lobes increases with the modal numbers, the overlap between the HG structure and the seed beam becomes progressively smaller for higher-order beams. This results in a decrease in the fidelity of the generated beam. 

Both the LG and HG modes display a dependence of the fidelity on the mode number of the constituent basis functions due to the radial dependence of the transverse field, which bottlenecks the mode capacity in diffraction-limited systems. Nevertheless, it is possible to address this by defining an orthogonal basis in which the spatial modes depend only on the azimuthal coordinate. One such implementation is a basis formed by the angular (ANG) modes, which is named after their angularly varying intensity distribution. The ANG modes are written as

\begin{gather}
\label{ANG}
    U_j(r)=\frac{1}{\sqrt{d}}\sum_{l=-L}^L \text{LG}^l_0 (\mathbf{r}) e^{-i2\pi \frac{jl}{d}},\\ L = \begin{cases}
d/2,\ \text{if}\ \textit{d}\ \text{is even} \\
(d-1)/2,\ \text{if}\ \textit{d}\ \text{is odd}
\end{cases} \nonumber
\end{gather}
where $d$ determines the number of bright lobes azimuthally distributed around the optical axis, and the modal number $j$ determines the azimuthal location of the intensity maxima. When $d$  is even, the term with $l=0$ is excluded from the sum in Eq. \ref{ANG} to ensure a balanced superposition of OAM modes. The ANG modes form a mutually unbiased basis with the OAM basis \cite{mirhosseini-2015}. We calculate the fidelity (Eq. \ref{fide}) of transferring an ANG mode from the pump to the stimulated field using

\begin{gather}
    F_d(\gamma_\textit{sp})=\nonumber\\\frac{\left(\sum_l\int_0^\infty e^{- \frac{u}{2\gamma_\textit{sp}^2}} e^{-u } u^{|l|} [L_0^{|l|} (u)]^2 du\right)^2}{\sum_{l,l'}\int_0^\infty e^{-u } u^{|l|} [L_0^{|l|} (u)]^2 du \int_0^\infty e^{- \frac{u}{\gamma_\textit{sp}^2}} e^{-u } u^{|l'|} [L_0^{|l'|} (u)]^2 du}
    \label{fidANG}
\end{gather}

\begin{figure}[ht]
    \centering
    \includegraphics[width=0.8\linewidth]{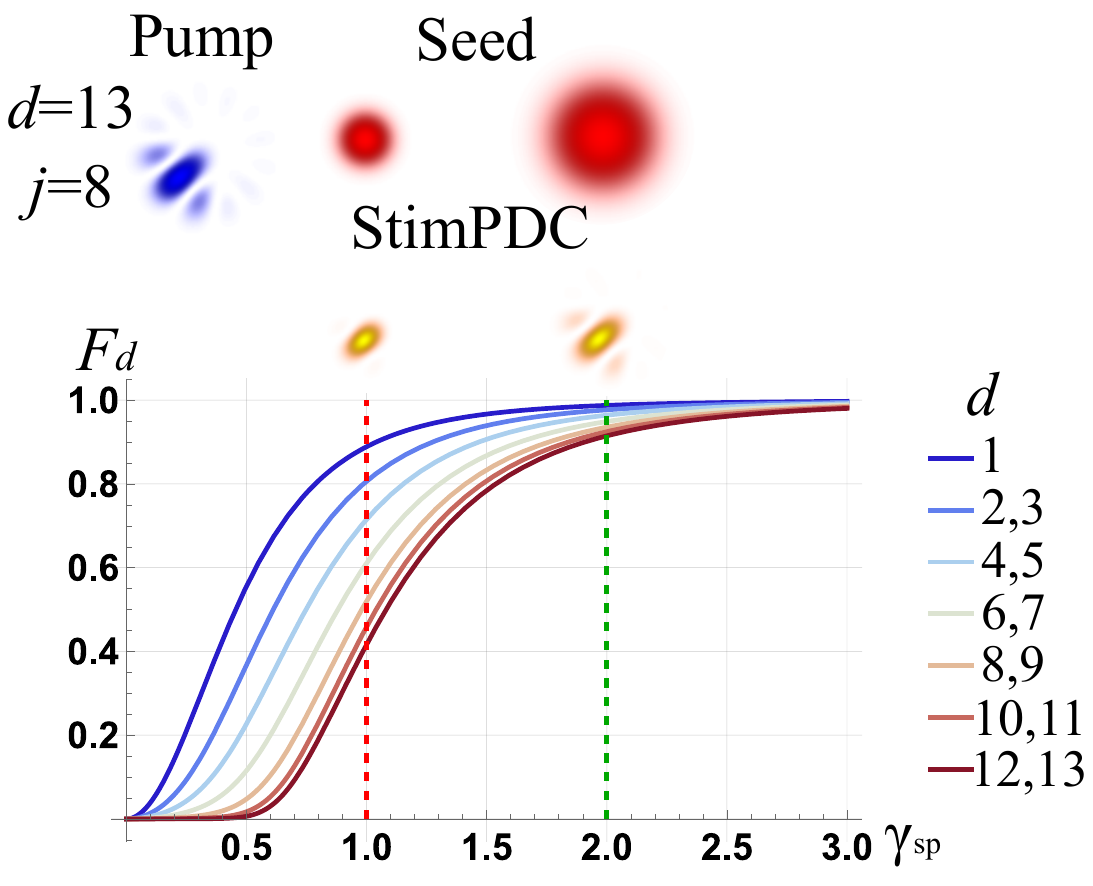}
    \caption{Fidelity plots of stimulated beams calculated using Eq. \ref{fidANG} for ANG beams with $ d\in[1,13]$ as function of $\gamma_\textit{sp}$. It is worth noting that fidelity decreases as the dimensionality increases. For even values of $d$, the $l=0$ component is excluded, resulting in modes with the same transverse area as those corresponding to the adjacent odd dimension $d+1$, which includes the $l=0$ term, because aside from this additional Gaussian component, both sets of modes are composed of the same $l$ contributions, which makes both of them have the same fidelity. The insets show examples of the transverse intensity profiles.}
    \label{fig:ang}
\end{figure}

Fig. \ref{fig:ang} shows the fidelity values for transferring ANG modes with different dimensions and modal numbers. Here, fidelity does not depend on the modal number \( j \), but rather solely on the dimensionality of the basis \( d \). The fidelity decreases as the dimensionality of the basis increases. This is because a larger dimensionality would require higher values of \( l \) in Eq. \ref{ANG}, which increases the size of the transverse spatial distribution for each basis element, thus decreasing the overlap between the seed beam and the pump ANG modes. However, for a fixed dimension, all the basis elements are composed of the same OAM modes, while what changes is the intermodal complex phase of the superposition. Therefore, the size of the transverse spatial distribution remains the same for different modal numbers. As a result, every element within the same basis will have homogenized fidelity as a function of $\gamma_\textit{sp}$. 

This analysis was carried out assuming that the spatial mode intended for transfer is embedded in the pump beam. This remains applicable in the scenario where the seed beam has a non-trivial complex amplitude profile while the pump beam is Gaussian beam. In this specific case, the stimulated beam corresponds to the complex conjugate of the seed beam’s spatial profile as seen in Eq. \ref{nolinear}.

\section{Experiments and discussion}
\label{experiments}

\subsection{Experimental setup}
\begin{figure}[ht]
\centering
    \includegraphics[width=1\linewidth]{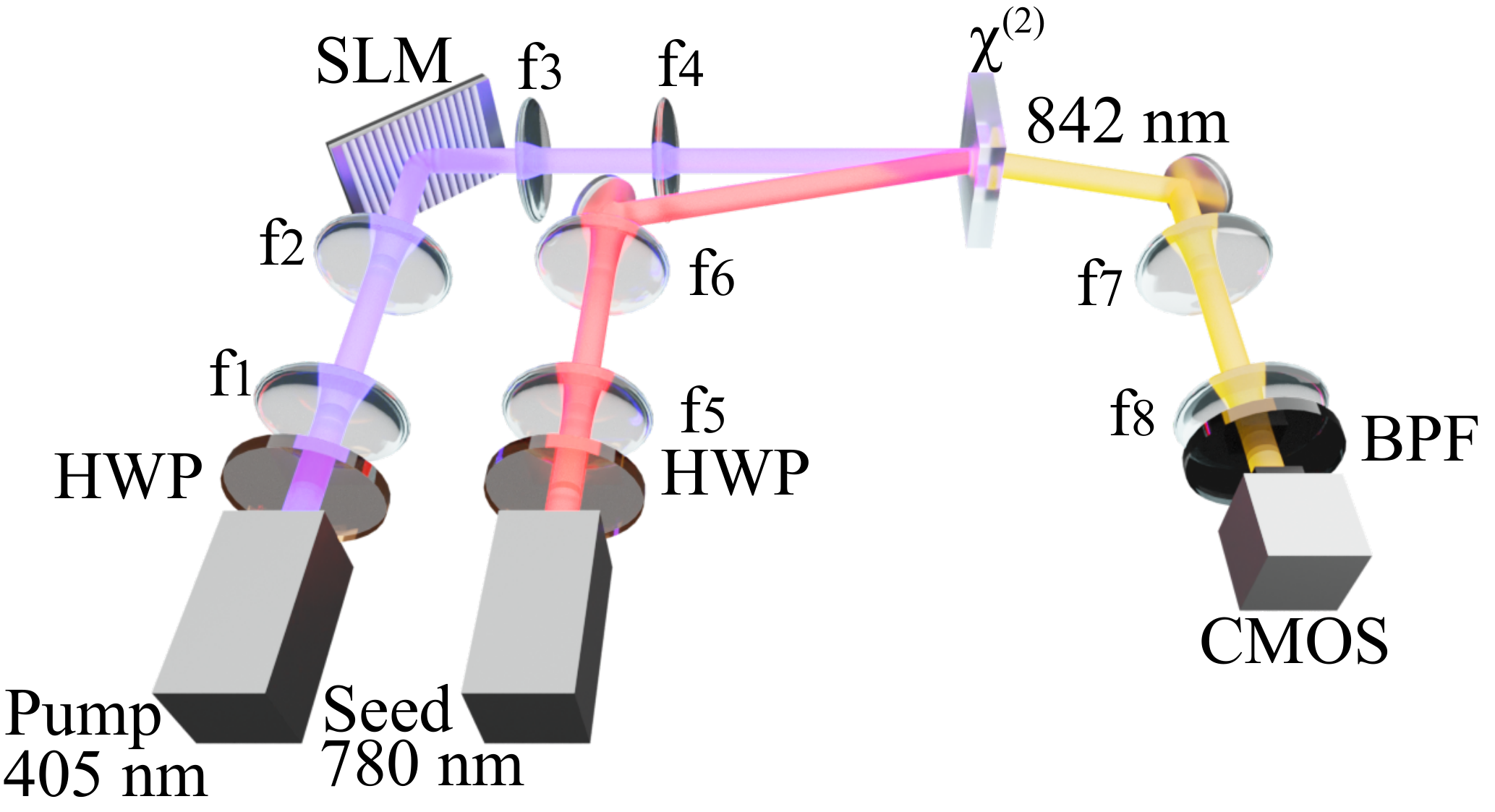}
    \caption{A collimated 405 nm beam is directed onto the SLM to generate the desired spatial modes, which are then imaged onto the nonlinear crystal $\chi^{(2)}$ to serve as the pump. A Gaussian beam at 780 nm is seeded to stimulate the PDC process, and its beam waist is varied. We achieve optimal phase-matching by tuning the polarizations input beams using HWPs. Finally, the stimulated beam is imaged onto a CMOS camera to capture its intensity. The CMOS camera is translated along the optical axis to record the beam intensity at different propagation distances.}
    \label{fig:experimento1}
\end{figure}

We experimentally validate our theoretical predictions using the setup depicted in Fig. \ref{fig:experimento1}. A continuous wave pump laser ($\lambda_p=$ 405 nm, power = 20 mW) is collimated using lenses \( f_1 \) = 250 mm and \( f_2 \) = 500 mm. A spatial light modulator (SLM) encodes the laser beam with a desired spatial structure. After SLM, the pump beam retains 2mW of power. The structured pump beam is then imaged with lenses \( f_3 \) = 500 mm and \( f_4 \)= 250 mm onto a 2-mm-long BBO crystal cut for type II phase-matching. The seed laser ($\lambda_s=$ 780 nm, power = 20 mW) is aligned at $4^\circ$ with respect to the pumping beam, and its beam waist at the crystal plane is adjusted by changing \( f_5 \) and \( f_6 \) to obtain different $\gamma_{sp}$ measurements. The polarizations of the beams are controlled using half-wave plates (HWPs) such that the two beams have orthogonal polarizations with respect to each other. The idler beam ($\lambda_i=$ 842 nm) is directed onto a CMOS camera (pixel size 3.45 $\mu$m $\times$ 3.45 $\mu$m) using two lenses with focal lengths \( f_7\) =\( f_8\)= 250 mm. The camera is equipped with a band-pass filter centered at 840 nm with a 20 nm bandwidth.

\subsection{Tailoring beam properties}

\begin{figure}[ht]
    \centering
    \includegraphics[width=0.8\linewidth]{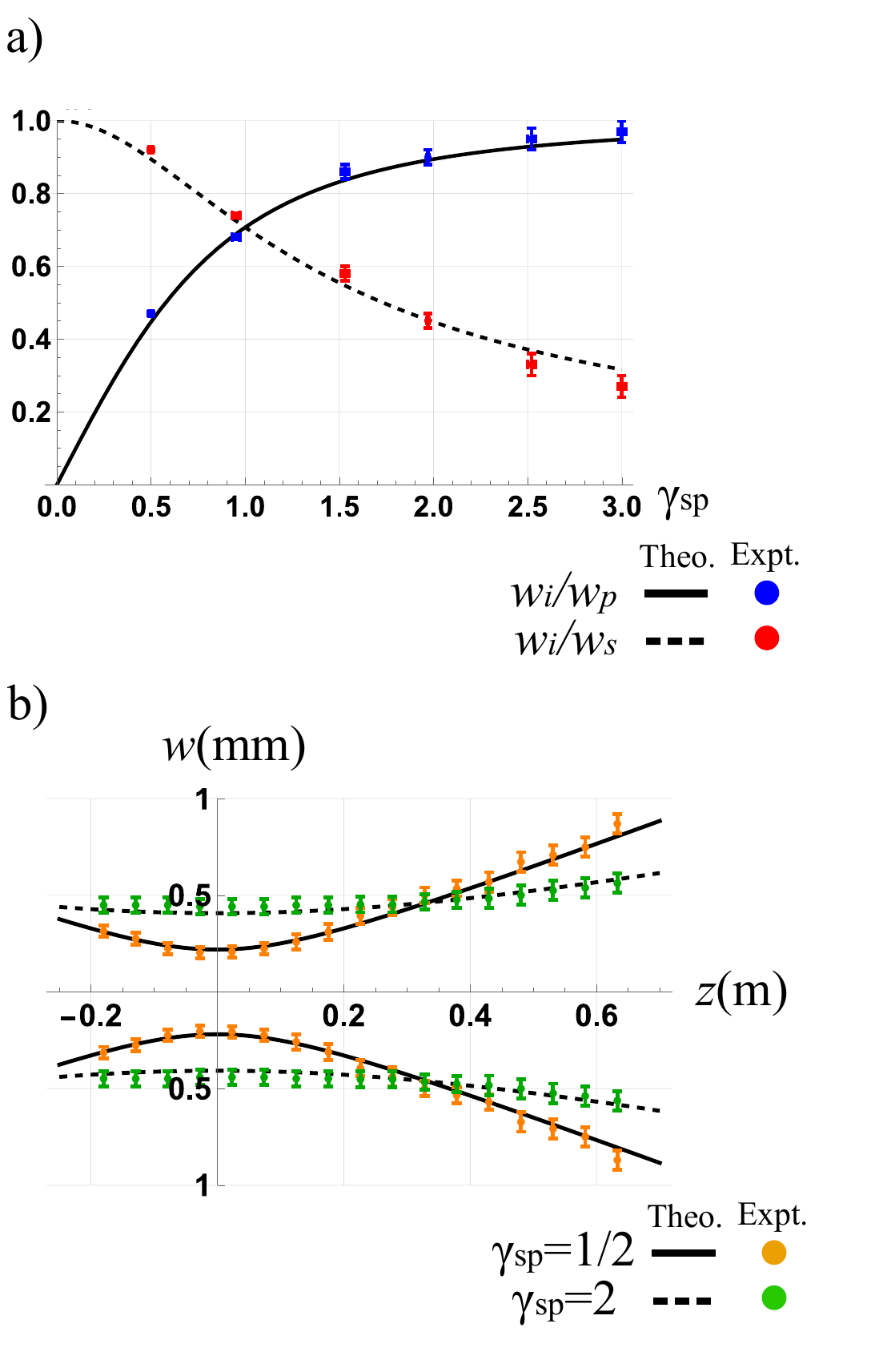}
    \caption{Theoretical and experimental comparison of: \textbf{a)} $w_i/w_p$ and $w_i/w_s$ given by Eqs. \ref{waist} . Here the pump and seed beams are Gaussian beams with waist at $w_{p}$= 300 $\mu$m and the seed beam waist $w_{s}$ is varied from $w_{s}$= 150 $\mu$m to $w_{s}$= 900 $\mu$m.  \textbf{b)} The beam waist profile over a propagation of 85 cm to show the tuning on the Rayleigh range of the stimulated beam given in Eq. \ref{range} . Here we change $\gamma_\textit{sp}$ by setting the seed beam waist at $w_{p}$= 300 $\mu$m (orange plot) and $w_{p}$= 600 $\mu$m (green plot). The error bars are given by the fit algorithm.}
    \label{fig:result1}
\end{figure}

As a first step, we demonstrate experimental control over the spatial features of the stimulated beam by varying $\gamma_{sp}$. The desired waist parameter of the pump beam is imposed by generating a fundamental Gaussian mode using the SLM. To estimate $w_p$, $w_s$, and $w_i$, we fit Eq. \ref{gauss} to the intensity profile captured in the crystal plane. By varying $\gamma_\textit{sp} = w_s/w_p$, we can compare experimentally measured $w_i/w_p$ and $w_i/w_s$ with Eqs. \ref{waistp} and \ref{waists}. We performed this procedure for $\gamma_\textit{sp}\in [0.5,3]$ with a 0.5 step size. These results are shown in Fig. \ref{fig:result1}.a. 

If the waists of the pump and seed beam are equal ($\gamma_\textit{sp}=1$), the stimulated beam will be $2^{-1/2}$ times the waist of the incoming beam, resulting in a difference in the waist of the beam of approximately 30\%, effectively truncating the spatial information of both input beams. However, if the beam waist of the seed increases compared to that of the pump ($\gamma_\textit{sp}> 1$), the energy interaction area between the pump and the seed beam will increase. As a result, the waist of the stimulated beam asymptotically converges to that of the pump beam. For instance, expanding the seed beam waist to twice the pump beam waist ($\gamma_\textit{sp}=2$) reduces the size difference to 10\%.

By shifting the camera along the optical axis in a range of 85 cm, we measure the evolution of the transverse profile during free-space propagation. By setting $w_{p}$= 300 $\mu$m and changing $\gamma_\textit{sp}$ from 1 to 2, we calculate that the Rayleigh range of the stimulated beam (Eq. \ref{range}) changes from 0.17 m to 1.06 m. Fig. \ref{fig:result1}b compares the experimentally measured beam widths at different propagation distances with $w(z)=w_0 \sqrt{1+\left(\frac{z}{z_\textit{R}}\right)^2}$ evaluated for the specified Rayleigh range values. Our results exhibit consistency between theory and experiment. 

\subsection{Improving fidelity of spatial modes}

\begin{figure}[ht]
    \centering
    \includegraphics[width=1\linewidth]{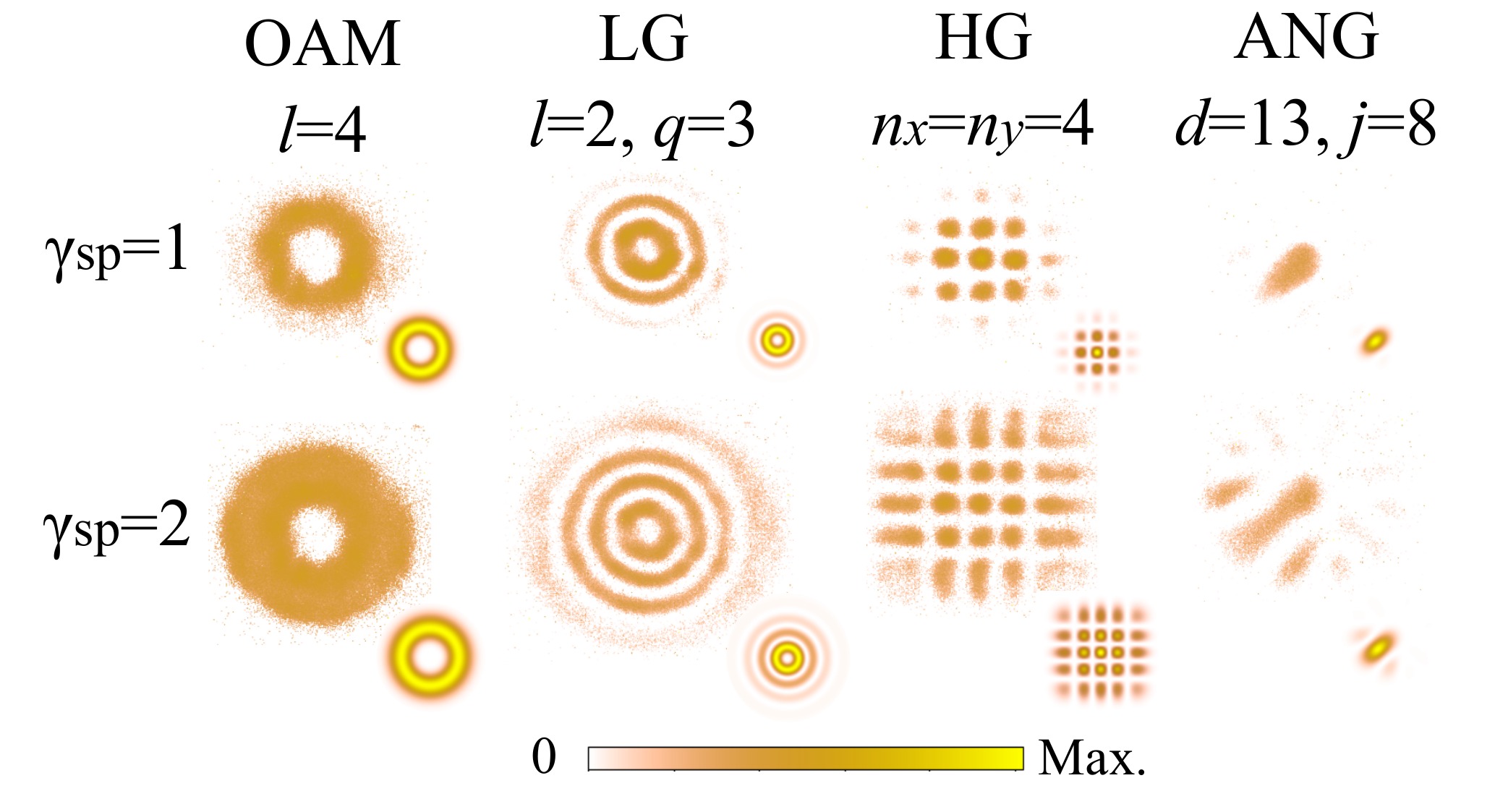}
    \caption{ Normalized intensity captures of the stimulated beam for $\gamma_\textit{sp}=1$ and $\gamma_\textit{sp}=2$, for different modes. The inset figures beneath each capture display the theoretical normalized intensity profiles derived in the preceding section.}
    \label{fig:ccd}
\end{figure}

\begin{figure}[ht]
    \centering
    \includegraphics[width=1\linewidth]{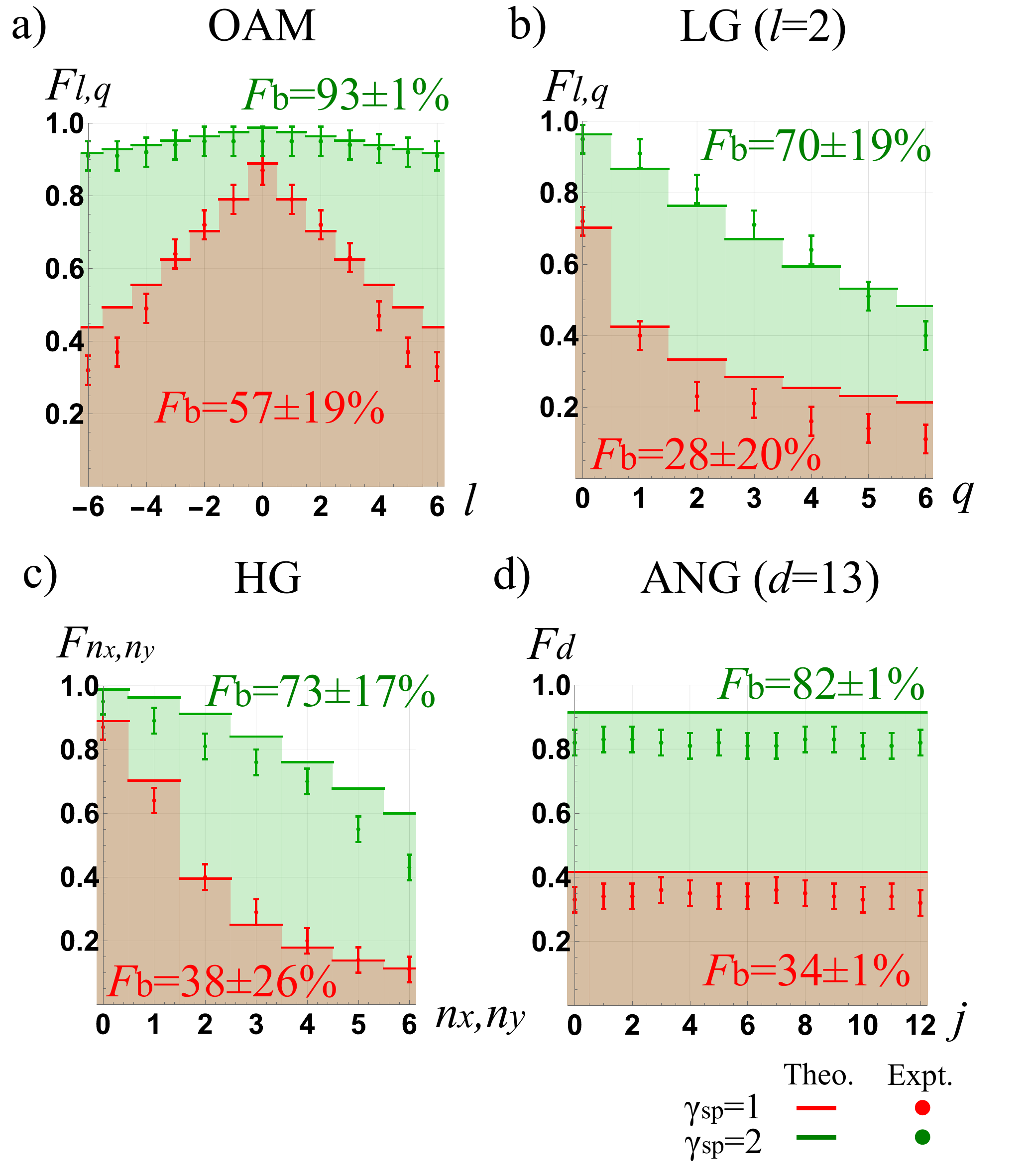}
    \caption{ Theoretical and experimental comparison of fidelity for the spatial modes: \textbf{a)} OAM ($q=0$) for $l\in[-6,6]$. \textbf{b)} LG with fixed $l=2$ and with $q\in[0,6]$. \textbf{c)} HG basis for $ n_x = n_y \in[0,6]$. \textbf{d)} ANG basis with $d=13$. The error bars are given by the signal to noise ratio of the CMOS captures.}
    \label{fig:results}
\end{figure}

In free-space optical communication, it is important to retain a near-uniform fidelity across the utilized spatial modes without resorting to beams with excessive size contrast or compromising the thin-crystal approximation. Here we identify the optimal choice of technical parameter (specifically, $\gamma_\text{sp}$) and spatial mode basis by comparing the corresponding fidelity values measured in experiments. Given the intensities of the pump and stimulated fields on the crystal plane, we can estimate the fidelity as: 
\begin{equation}
    F=\frac{\left(\sum_{ x,y} \sqrt{ I_{p}(x,y) I_{i}(x,y)} \right)^2}{\sum_{ x,y} I_p(x,y)\sum_{ x,y} I_{i}(x,y)},
\end{equation}
where $I_{p,i}(x,y)=|U_{p,i}(x,y)|^2$. Fig. \ref{fig:ccd} shows examples of experimental normalized intensity captures for different modes, and Fig. \ref{fig:results} shows the average fidelity values, expressed as percentages, obtained for each mode bases.

For LG and HG bases, decreasing $\gamma_\textit{sp}$ causes the transfer fidelity to reduce more at higher mode orders. This effect could reduce the effective spatial dimensionality and restrict the information capacity of StimPDC devices. As shown in Fig. \ref{fig:results}a, with a decreased $\gamma_\textit{sp}$ from 2 to 1, LG modes with dimensionality $d=13$ ($q=0$ and $l\in[-6,6]$) would display a reduced average fidelity with an increased variation, from $93\pm 1\%$ to $57\pm 19\%$. Similar results are seen for beams in LG modes with different radial indices ($l=2$ and $q\in[0,6]$) and HG modes with different mode indices ($n_x=n_y\in[0,6]$), as depicted in  Fig. \ref{fig:results}b and c, respectively. Although ANG modes experience a reduced average fidelity from $82\%$ to $34\%$, as shown in Fig. \ref{fig:results}d, they retain a fidelity consistency of $\pm 1\%$ across all mode orders. The discrepancy between the experimental and theoretical fidelities of the ANG modes may be attributed to imperfections in the phase masks used for their generation. Nevertheless, we show that when multiple modes are employed, ANG modes exhibit greater consistency across the basis, whereas LG and HG modes display mode-dependent variations in fidelity. This feature makes ANG modes particularly advantageous for applications such as free-space quantum key distribution \cite{mafu-2013,scarfe-2025}, where encoding and transmitting information evenly across spatial modes is crucial. 

With $\gamma_\textit{sp}$ increased from 1 to 2, the average mode fidelity in all bases improved. This result agrees with our theoretical predictions and has significant implications for practical applications using StimPDC. Since maintaining information transfer fidelity via StimPDC requires increasingly large $\gamma_{sp}$ at higher mode orders, one needs to either make the seed beam waist increasingly large or make the pump beam waist diminishingly small. However, the commercially available nonlinear crystals typically have clear apertures of a few millimeters, making it impractical to use large seed beams without inducing diffraction effects. On the other hand, reducing the beam waist of the pump decreases its Rayleigh length and could invalidate the thin crystal approximation. In this case, the nonlinear propagation effects could become prominent and introduce distortion to the stimulated beam.

\section{Conclusions}
\label{section5}

We present a comprehensive study on the transfer of transverse spatial structure and longitudinal propagation properties in StimPDC. Specifically, we theoretically and experimentally investigate the dependence of the Rayleigh range and mode fidelity of the stimulated beam on the wavelength, beam waist, and Rayleigh range of the pump and seed beams. We demonstrate this dependence by showcasing the transfer of LG, HG and ANG modes in StimPDC. Our results show that while the transfer fidelity of ANG modes depends solely on the ratio of the beam waists of both input beams ($\gamma_\textit{sp}=\frac{w_s}{w_p}$), that of LG and HG modes is also influenced by the mode order. Therefore, the fidelity of mode transfer in StimPDC can be optimized by both tuning the input beam parameters and choosing an ideal spatial mode basis. Following this principle, we successfully transferred complex spatial structures in LG, HG and ANG modes from the pump beam to the stimulated output beam with high fidelities.

Our work provides technical guidance for controlling structured light using nonlinear optical methods and has significant implications for implementing StimPDC for free-space optical communication applications. For instance, since the ANG basis maintains a nearly constant fidelity across all mode orders, it could be preferred for ensuring inter-modal information consistency. Although our experiments were conducted using fixed wavelengths, our theoretical model applies to StimPDC using pump and seed beams with arbitrary phase-matched wavelengths. This capability could enable control of the spatial structures of optical beams with spectral tuning. These findings advance structured light manipulation in nonlinear optics and underscore the potential of StimPDC for high-fidelity mode transfer in optical communication systems.

\nocite{*}


\begin{thebibliography}{23}%
\makeatletter

\bibitem{BOUCHARD-2018}\href{https://doi.org/10.22331/q-2018-12-04-111}{F. Bouchard, K. Heshami, D. England, R. Fickler,R.W. Boyd, B. Englert,L.L. Sánchez-Soto, and E. Karimi,  Quantum, \textbf{2}, 111. (2018)}
\bibitem{klug-2023}\href{https://doi.org/10.1117/1.AP.5.1.016006}{A. Klug,  C. Peters, and A. Forbes,, Adv. Photonics \textbf{5} (2023).}
\bibitem{boyd:20}\href{https://doi.org/10.1364/OE.419493}{  K. K. M.Bearne,Y. Zhou, B. Braverman, J. Yang, S.A. Wadood, A.N. Jordan, A.N. Vamivakas, Z. Shi, and R.W. Boyd, Optics Express, \textbf{29}(8), 11784 (2021).}
\bibitem{Fang:24} \href{https://doi.org/10.1038/s41377-024-01386-5}{X. Fang, X. Hu, B. Li, H. Su, K. Cheng, H. Luan, and M. Gu, Light Science and Applications, \textbf{13}(1) (2024).}
\bibitem{rosales2017shape}\href{https://doi.org/10.1117/3.100024}{C. Rosales-Guzmán and A. Forbes, Society of Photo-Optical Instrumentation Engineers (SPIE) (2017).}
\bibitem{rubano2019q} \href{https://doi.org/10.1364/JOSAB.36.000D70}{A. Rubano, F. Cardano, B. Piccirillo, and L. Marrucci, J. optical society America B \textbf{36}, D70–D87 (2019).}
\bibitem{dorrah2022tunable}\href{10.1126/science.abi6860}{ A.H. Dorrah and F. Capasso, Science, \textbf{376}(6591) (2022).}
\bibitem{buono-2022} \href{10.29026/oea.2022.210174}{W.T. Buono and A. Forbes, Opto-Electronic Adv. \textbf{5}, 210174 (2022).}
\bibitem{jones2024ultrafast}\href{https://doi.org/10.1364/OL.531092}{ W. M. Jones and M. A. Reber, Opt. Lett. \textbf{49}, 4999–5002 (2024).}
\bibitem{de-oliveira-2019} \href{https://pubs.acs.org/doi/10.1021/acsphotonics.9b01524}{A.G. de Oliveira, M. F. Z. Arruda, W. C. Soares, S. P. Walborn, R. M. Gomes, R. M. de Araújo, P. H. S. Ribeiro, ACSPhotonics \textbf{7}, 249–255 (2019).}
\bibitem{dos2022phase} \href{https://doi.org/10.1515/nanoph-2021-0502}{G.H. dos Santos , A.G. de Oliveira , N. R. da Silva, G. Cañas , E. S. Gómez , S. Joshi , Y. Ismail , P. H. S. Ribeiro and S. P. Walborn, Nanophotonics \textbf{11}, 763–770 (2022).}
\bibitem{Liscidini:13} \href{https://doi.org/10.1103/PhysRevLett.111.193602}{M. Liscidini and J. E. Sipe, Phys. Rev. Lett. \textbf{111}, 193602 (2013).}
\bibitem{fadrny-2024}\href{https://doi.org/10.1038/s41534-024-00885-y}{J. Fadrný, M. Neset, M. Bielak, M. Ježek, J. Bílek and J. Fiurášek ,  npj Quantum Inf. \textbf{10} (2024).}
\bibitem{xu-2024}\href{https://doi.org/10.1103/PhysRevResearch.6.L042047}{Y. Xu, S. Choudhary, and R.W. Boyd, Phys. Rev. Res. \textbf{6} (2024).}
\bibitem{Xu:23} \href{https://doi.org/10.1364/OE.506383}{Y.Xu, S.Tang, A.N.Black, and R.W.Boyd, Opt. Express \textbf{31}, 42723–42729 (2023).}
\bibitem{de-oliveira-2020} \href{https://doi.org/10.1103/PhysRevApplied.14.024048}{A. de Oliveira, N. R. da Silva, R. M. de Araújo, P.H. Souto Ribeiro, and S.P. Walborn, Phys. Rev. Appl. \textbf{14} (2020).}
\bibitem{singh-2025} \href{https://doi.org/10.1364/OE.562028}{S. Singh, I. Nape, and A. Forbes, Opt. Express \textbf{33}, 27615-27625 (2025).}
\bibitem{arruda-2018} \href{https://doi.org/10.1103/PhysRevA.98.023850}{M. F. Z. Arruda, W. C. Soares, S. P. Walborn, D.S. Tasca, A. Kanaan, R.M. de Araújo, and P. H.S. Ribeiro, Phys. review. A \textbf{98} (2018).}
\bibitem{Boyd:2020} \href{https://www.sciencedirect.com/book/9780123694706/nonlinear-optics#book-description}{R. W. Boyd, Academic Press, 2020.}
\bibitem{saleh-1991} \href{https://onlinelibrary.wiley.com/doi/book/10.1002/0471213748#aboutBook-pane}{B. E. A. Saleh and M. C. Teich (1991).}
\bibitem{mirhosseini-2015} \href{https://iopscience.iop.org/article/10.1088/1367-2630/17/3/033033}{M. Mirhosseini, O. S. Magaña-Loaiza, M. N. O’Sullivan, B. Rodenburg, M. Malik, M. P. J. Lavery, M.J. Padgett, D.J. Gauthier and R.W. Boyd, New J. Phys. \textbf{17}, 033033 (2015).}
\bibitem{mafu-2013}\href{https://doi.org/10.1103/PhysRevA.88.032305}{M. Mafu, A.Dudley, S.Goyal, D. Giovannini, M. McLaren, M.J. Padgett, T. Konrad, F. Petruccione, N. Lutkenhaus, and A. Forbes, Phys. Rev. A \textbf{88} (2013).}
\bibitem{scarfe-2025}\href{https://doi.org/10.1038/s42005-025-01986-6}{ L.Scarfe, F. Hufnagel, M. F. Ferrer-Garcia,  A. D’Errico, K. Heshami and Ebrahim Karimi, Commun. Phys. \textbf{8} (2025).}
\end{thebibliography}
%

\end{document}